\documentstyle[12pt]{article}

\newcommand{\be}{\begin{equation}}
\newcommand{\ee}{\end{equation}}
\newcommand{\bea}{\begin{eqnarray}}
\newcommand{\eea}{\end{eqnarray}}
\newcommand{\ba}{\begin{array}}
\newcommand{\ea}{\end{array}}
\newcommand{\vs}[1]{\vspace{#1 mm}}

\def\bbox{{\,\lower0.9pt\vbox{\hrule \hbox{\vrule height 0.2 cm
\hskip 0.2 cm \vrule height 0.2 cm}\hrule}\,}}
\newcommand{\dsl}{\pa \kern-0.5em /}

\newcommand{\pa}{\partial}

\font\mybb=msbm10 at 12pt
\def\bb#1{\hbox{\mybb#1}}
\def\bZ {\bb{Z}}

\def\ua{{\underline a}}
\def\ub{{\underline b}}

\def\CF {{\cal F}}

\begin{document}

\topmargin 0pt
\oddsidemargin 5mm

\renewcommand{\thefootnote}{\fnsymbol{footnote}}
\begin{titlepage}

\setcounter{page}{0}
\begin{flushright}
UG-4/98\\
hep-th/9804011
\end{flushright}

\vs{5}
\begin{center}
{\Large SUPER D-BRANES REVISITED}
\vs{10}

{\large
E. Bergshoeff$^1$ and  P.K. Townsend$^2$\footnote{On
leave from DAMTP,  University of Cambridge, U.K.}
} \\
\vs{5}
${}^1${\em Institute for Theoretical Physics, Nijenborgh 4,\\
9747 AG Groningen, The Netherlands.}\\
\vs{5}
${}^2${\em Institute for Theoretical Physics,\\
University of California at Santa Barbara,\\ 
CA 93106, USA.}
\end{center}
\vs{10}
\centerline{{\bf Abstract}}

A version of the $\kappa$-symmetric super D-p-brane action
is presented  in which the tension is a dynamical variable, equal to the flux of
a p-form worldvolume gauge field. The Lagrangian is shown to be invariant under
all (super)isometries of the background for appropriate transformations of
the worldvolume gauge fields, which determine the central charges in the
symmetry algebra. We also present the hamiltonian form of the action
in a general supergravity background.

\end{titlepage}
\newpage
\renewcommand{\thefootnote}{\arabic{footnote}}
\setcounter{footnote}{0} 

\section{Introduction}

It is now widely appreciated that super p-branes in a vacuum spacetime
background are associated with p-form extensions of the standard spacetime
supersymmetry algebra. In the usual formulation, and for the branes of the
`old branescan', this extension arises as a consequence of the non-invariance
under spacetime supersymmetry transformations of a Wess-Zumino (WZ)
term in the worldvolume Lagrangian \cite{azc}. The p-form is proportional
to the tension $T$ and can be considered as a type of `classical anomaly' in
which the role of Planck's constant is taken by $T$ \cite{azcb}. A simple
example is the D=11 supermembrane, in which the WZ term is the pull-back
to the worldvolume of the 3-form superspace gauge potential of D=11
supergravity \cite{bst}.

This is not the full story, however. It was shown in \cite{tension,blt}
that the introduction of a p-form gauge potential $A$ allows the construction of
an alternative local worldvolume Lagrangian {\sl without} a WZ term. The new
worldvolume field does not lead to any additional local degrees of freedom
but does lead to an integration constant, which can be identified as the
p-brane tension T. When $T$ is non-zero the supersymmetry algebra must include
a p-form charge, as before, but since the new Lagrangian is invariant under
supersymmetry transformations this charge now has a different origin. To
explain this point, let us call the algebra deduced from commutators of
supersymmetry transformations acting  on worldvolume fields the `naive'
supersymmetry algebra. This need not be the same as the anticommutator
of Noether supercharges because central charges appearing in the latter may
commute with all worldvolume fields. For example, in the usual formulation of
the supermembrane, with WZ term, the worldvolume fields are the maps from the
worldvolume to superspace. The `naive' supersymmetry algebra in this case is
therefore the algebra of Killing vector superfields. Assuming a vacuum
superspace background, this is just the standard D=11 supersymmetry algebra,
which fails to include the 2-form extension implied by the presence of the WZ
term.

In the new formulation of the supermembrane there is no WZ term so the naive
supersymmetry algebra must include the 2-form extension. What makes
this possible is a non-trivial supersymmetry transformation of the 2-form
$A$, which is such that the commutator of two supersymmetry transformations
yields a gauge transformation with a parameter that determines the central
charge structure. This phenomenon can occur whenever the worldvolume fields
include gauge fields. It was first noted \cite{sorokin} in the context of the
2-form gauge potential appearing in the D=11 superfivebrane action of 
\cite{bandos}. A detailed verification in a very general context, including Type
II super D-branes \cite{swedes,bergstown,agan}, has since been provided by
Hammer \cite{hammer}. In this case the `naive' algebra includes a
Neveu-Schwarz/Neveu-Schwarz (NS) 1-form charge arising from the presence of the
Born-Infeld (BI) gauge field, while all Ramond/Ramond (RR) charges arise from
the WZ term. 

From the fully non-perturbative point of view there is no real distinction
between NS and RR charges, so one might expect there to exist an alternative,
manifestly supersymmetric, super D-brane action in which all p-form charges
appear already in the `naive' supersymmetry algebra. Here we construct this
action via the introduction of a p-form worldvolume gauge potential, as
suggested by the supermembrane example. This new, manifestly supersymmetric,
formulation of the super D-p-brane can be motivated in a number of other ways.
It was pointed out in \cite{pktb} that boundaries (on M5-branes or on
`M-9-branes') of open supermembranes  could be interpreted as discontinuities in
the field strength of a 2-form potential on the D=11 supermembrane worldvolume,
and that M-theory and superstring dualities would then imply the existence of
p-form gauge potentials on the worldvolumes of most other M-theory and
superstring p-branes, all D-branes in particular. The new super D-string action
turns out to be not only manifestly supersymmetric (in a vacuum background) but
also manifestly invariant under the $Sl(2;\bZ)$ duality of IIB superstring
theory \cite{martin}, and it seems to be a general feature that manifest
$Sl(2;\bZ)$ invariance of IIB D-p-brane actions requires the introduction of
worldvolume p-form gauge potentials \cite{ceder}. More recently, worldvolume
p-form gauge potentials have found to be an essential ingredient in the
formulation of `massive' p-brane actions \cite{BLO}, i.e. IIA branes in
a massive IIA supergravity background.

One aim of this paper is to update our previous work on super D-branes
\cite{bergstown} in light of the above discussion. Our results on p-form gauge
potentials are complementary to those of \cite{martin,ceder,BLO}. We do not
consider the implementation of $Sl(2;\bZ)$ invariance in the IIB case, nor
massive backgrounds in the IIA case. However, by virtue of these limitations we
are  able to discuss all D-branes at once, thereby highlighting their common 
features. The bosonic Lagrangian of the new super D-brane action is
simply\footnote{On setting $G$ to zero this reduces to the null D-brane
Lagrangian of \cite{lindstrom}.}
\begin{equation}
L= {1\over2v}[e^{-2\phi}\det(g+{\cal F}) + (\star G_{(p+1)})^2]
\label{onea}
\end{equation}
where $\phi$ is the dilaton, $v$ is an independent worldvolume density, $g$ is
the induced metric, and
\begin{equation}
{\cal F} = dV-B
\label{oneb}
\end{equation}
is the two-form field strength of the BI 1-form gauge
potential $V$, with $B$ the pullback to the worldvolume of the NS two-form gauge
potential. The scalar density $\star G_{(p+1)}$ is the worldvolume Hodge
dual of a (p+1)-form field strength $G_{(p+1)}$ for the new p-form worldvolume
gauge potential $A_{(p)}$. In order to deal with all D-p-branes simultaneously 
it is convenient to combine these p-form gauge potentials into the formal
sum\footnote{This has also been used in \cite{BLO}.}
\begin{equation}
A = \sum_{k=0}^9 A_{(k)}\, .
\label{onec}
\end{equation}
The field strength of $A$ is 
\begin{equation}
G= dA - C e^{\cal F}
\label{oned}
\end{equation}
where C is the formal sum of R-R gauge potentials introduced in \cite{douglas}.

The supersymmetric extension of the Lagrangian (\ref{onea}) is formally
identical, but with an N=2 superspace replacing the D=10 background spacetime.
What we have to show is that the Lagrangian so-obtained is $\kappa$-symmetric.
We do so in the following section.

The super-Poincar{\'e} invariance of the super D-brane action in a flat
superspace background is a special case of invariance under the
(super)isometries of any background allowed by $\kappa$-symmetry. Another aim
of this paper is to prove this statement. In general, the invariance requires an
appropriate choice of transformation rules for worldvolume gauge fields. If
these include the new p-form gauge field then the
Lagrangian (rather than just the action) is invariant. As mentioned above in the
context of supertranslations of flat superspace, the transformations of the
worldvolume gauge fields then determine the central charge structure of the
(super)algebra of isometries. Thus, the discussion above generalises directly to
an arbitrary background. This observation is especially important for the D3
brane since it implies that the super D3-brane action in the near-horizon
geometry of the D3-brane solution is invariant under the full $SU(2,2|4)$
isometry of that background. This can be interpreted as a non-linearly realized
superconformal symmetry of the super D3-brane action in this background,
generalizing the well-known superconformal symmetry of the free N=4 D=4
super-Maxwell action and the more-recently established \cite{malda,kaletal}
non-linearly-realized conformal invariance of the bosonic action.  

Finally, we present the hamiltonian formulation of our super D-brane action,
generalizing the bosonic results of \cite{kallosh,lindstrom} and the flat
background IIB results of \cite{kam}.

\section{$\kappa$-symmetry}

As in \cite{bergstown}, we define $\delta E^A \equiv \delta Z^M E_M{}^A$,
and set $\delta_\kappa E^a=0$, temporarily leaving open the choice of
$\delta_\kappa E^\alpha$. The $\kappa$-symmetry variation of the BI field $V$ is 
\begin{equation}
\delta_\kappa V_i = E_i{}^A \delta_\kappa E^\beta B_{\beta A}\, ,
\label{twoa}
\end{equation}
which is such as to ensure the `supercovariance' of the variation of ${\cal F}$.
Using the standard type II D=10 supergravity superspace constraints, summarized
in \cite{bergstown}, we find that
\begin{equation}
\delta_\kappa \det (g+{\cal F}) = -2i\delta_\kappa \bar E\tilde N^i E_i
\label{twob}
\end{equation}
where
\begin{equation}
\tilde N^i = \det (g+{\cal F})\, (g+{\cal F})^{ij}{\cal P}_+ \gamma_j +
\det (g-{\cal F})\, (g-{\cal F})^{ij}{\cal P}_- \gamma_j\, .
\label{twoc}
\end{equation}
Note that, because of the determinant, $\tilde N^i$ is non-singular even when
$(g \pm {\cal F})$ has no inverse. The (reducible) worldvolume Dirac matrices
$\gamma_i$ are the pullbacks $E_i{}^a\Gamma_a$ of the spacetime Dirac matrices
(more generally, the antisymmetrized product of $p$ worldvolume Dirac matrices
defines a matrix-valued worldvolume $p$-form $\gamma_{(p)}$ with components
$\gamma_{i_1\dots i_p}$), and
\begin{equation}
{\cal P}_\pm = \cases{{1\over2}(1 \pm \Gamma_{11}) & IIA \cr {1\over2}(1\pm
\sigma_3) & IIB.}
\label{twod}
\end{equation}
We find, similarly, that
\begin{equation}
\delta_\kappa [Ce^{\cal F}] = d(i_{\delta Z} C e^{\cal F}) +
i\delta_\kappa \bar E \gamma e^{\cal F} E
\label{twoe}
\end{equation}
where $i_{\delta Z}C$ is the contraction of $C$ with the vector superfield
$\delta_\kappa Z^M\partial_M$, and $\gamma$ is the formal sum
\begin{equation}
\gamma = \cases{ e^{-\phi} \sum_{p=0}^8 \gamma_{(p)} 
(\Gamma_{11})^{(p-2)(p-6)/4} & IIA \cr
e^{-\phi} \sum_{p=1}^9 \gamma_{(p)} {\cal P} (\sigma_1)^{(p-3)(p-7)/4}
(i\sigma_2)^{(p-1)(p-9)/4} & IIB }
\label{twof}
\end{equation}
where ${\cal P}$ is the IIB chiral projector on D=10 spinors.

We choose the $\kappa$-transformation of $A$ such as to ensure the
`supercovariance' of the transformation of the field strength $G$. This
requires 
\begin{equation}
\delta A= (i_{\delta Z} C) e^{\cal F}\, ,
\label{twog}
\end{equation}
which leads to
\begin{equation}
\delta_\kappa G = -i\delta_\kappa \bar E \gamma e^{\cal F} E\, .
\label{twoh}
\end{equation}
For a D-p-brane we must select the (p+1)-form in this formal sum, and then take
the worldvolume Hodge dual. At this point it is convenient to introduce the
matrix\footnote{The matrix $\Xi_{(0)}$, which is just $\sqrt{-\det g}$ times the
matrix $\Gamma_{(0)}$ of \cite{bergstown}, is well-defined even for a degenerate
worldvolume metric.}
\begin{equation}
\Xi_{(0)} = {1\over (p+1)!}\,
\varepsilon^{i_1\dots i_{p+1}}\gamma_{i_1\dots i_{p+1}}\, ,
\label{twoi}
\end{equation}
which satisfies
\begin{equation}
\Xi_{(0)}^2= (-1)^{p(p+1)/2}\; \det g\ .
\label{twoj}
\end{equation}  
We may then write the variation of the worldvolume scalar $\star G$ as
\begin{equation}
\delta_\kappa \star G = -ie^{-\phi}\delta_\kappa \bar E \tilde M^i_{(p)} E_i
\label{twok}
\end{equation}
where\footnote{We have absorbed some factors into the definition of this matrix
relative to the matrix $M^i_{(p)}$ of \cite{bergstown}, where they instead appear
in eq. (4.13). The subsequent calculations of \cite{bergstown}, to be summarized
below, actually refer to a matrix $M^i_{(p)}$ {\it with} these factors, as is
clear  from the definition of the matrix $K^i_{(p)}$ in \cite{bergstown}.}
\begin{equation}
\tilde M_{(p)}^i = \sum_{n=0} {1\over 2^n n!} \gamma^{ij_1k_1\dots j_nk_n}\,
\Xi_0\, {\cal F}_{j_1k_1}\dots {\cal F}_{j_nk_n} \times \cases{
(-\Gamma_{11})^{n+ (p-2)/2} & IIA \cr (-\sigma_3)^{n+(p-3)/2} \; i\sigma_2 & IIB}
\label{twol}
\end{equation}

We may now put together the above results to find the $\kappa$-variation
of the proposed new super D-brane Lagrangian, at least in bosonic
backgrounds. In such backgrounds the dilaton has vanishing $\kappa$-variation,
so we have
\begin{equation}
\delta_\kappa [e^{-2\phi}\det(g+{\cal F}) + (\star G)^2]=
-2ie^{-2\phi}\delta_\kappa \bar E \big(\tilde N^i + e^\phi (\star G)\tilde
M^i_{(p)}\big) E_i\ .
\label{twom}
\end{equation} 
Given the results in \cite{blt,bergstown}, we see that the spinor variation
$\delta_\kappa E$ must take the form
\begin{equation}
\delta_\kappa \bar E_\alpha = 
\big[\bar\kappa (e^\phi \star G + \Xi)\big]_\alpha
\label{twon}
\end{equation}
where $\Xi$ is a matrix with the property
\begin{equation}
\Xi^2 = -\det (g+{\cal F})\ .
\label{twoo}
\end{equation}
Clearly $\Xi$ must reduce (up to a sign) to $\Xi_0$ when ${\cal F}$ vanishes.
Again using the results of \cite{bergstown} it follows that 
\begin{equation}
\Xi = \sum_{n=0}^\infty {1\over 2^n n!} 
\gamma^{j_1k_1\dots j_nk_n} {\cal F}_{j_1k_1}\dots {\cal F}_{j_nk_n}{\cal
J}^{(n)}_{(p)}
\label{twop}
\end{equation}
where 
\begin{equation}
{\cal J}^{(n)}_{(p)} = \cases{(\Gamma_{11})^{n+(p-2)/2}\, \Xi_{(0)} & (IIA)\cr
(-1)^n (\sigma_3)^{n+(p-3)/2}\; i\sigma_2 \otimes \Xi_{(0)} & (IIB)}
\label{twoq}
\end{equation}

We now claim that
\begin{equation}
e^{-2\phi}(e^\phi\star G + \Xi)\big(\tilde N^i + 
e^\phi (\star G)\tilde M^i_{(p)}\big)
\equiv \tilde M^i_{(p)} [e^{-2\phi}\det(g+{\cal F}) + (\star G)^2]\, .
\label{twor}
\end{equation}
The basis for this claim is the calculation sketched in \cite{bergstown} in the
course of which it was noted that terms involving a square root of a
determinant cancel separately. Thus, the calculation of \cite{bergstown} actually
establishes {\sl two} identities, which are
\begin{equation}
\tilde N^i + \Xi \tilde M^i_{(p)} \equiv 0, \qquad \Xi \tilde N^i - \det (g+
{\cal F})
\tilde M^i_{(p)} =0\, .
\label{twos}
\end{equation}
These identities are precisely those needed for (\ref{twor}), from which 
it follows that
\begin{equation}
\delta_\kappa [e^{-2\phi}\det(g+{\cal F}) + (\star G)^2] = -2i(\bar\kappa \tilde
M^i_{(p)} E_i) [e^{-2\phi}\det(g+{\cal F}) + (\star G)^2]\, .
\label{twot}
\end{equation}
The new super D-p-brane Lagrangian is therefore invariant provided that we 
choose the $\kappa$-variation of the Lagrange multiplier to be
\begin{equation}
\delta_\kappa v = -2iv(\bar\kappa \tilde M^i_{(p)} E_i)\, .
\label{twou}
\end{equation}
We have now established $\kappa$-symmetry. The $A$ field equation implies that
$\star G$ is a constant. The remaining equations are then those of the standard
super D-brane action with the tension equal to the constant $\star G$.

\section{Symmetries from background isometries}

We shall now consider symmetries of the new super D-brane action arising from
isometries of the background. These are generated by Killing vector superfields 
with respect to which the Lie derivatives of all superspace field strengths
vanish. This implies, in particular that the induced metric $g_{ij}$ is
invariant. Let $\xi_\alpha = \xi_\alpha^M(Z)\partial_M$ be the set of Killing
vector superfields with (anti)commutators
\begin{equation}
[\xi_\beta, \xi_\gamma] = 
f_{\beta\gamma}{}^\alpha \xi_\alpha \, ,
\label{syma}
\end{equation}
so that $f_{\beta\gamma}{}^\alpha$ are the structure constants of the Lie
(super)algebra of isometries. It will prove convenient to introduce the
transformations on spacetime superfields generated by $\xi_\alpha$ via a
BRST operator $s$. Thus,
\begin{equation}
sZ^M = c^\alpha\xi_\alpha^M 
\label{symb}
\end{equation}
where $c^\alpha$ are set of {\it constant} ghost `fields' with BRST
transformation
\begin{equation}
sc^\alpha = {1\over2} c^\gamma c^\beta f_{\beta\gamma}{}^\alpha\, .
\label{symc}
\end{equation}
Note that $s^2c^\alpha$ is identically zero as a consequence of the
Jacobi identity for the structure constants, and that the action of $s^2$ on
$Z^M$, and hence on all superfields, also vanishes (this being equivalent to
$sc^M\equiv 0$). 

Since $H=dB$ and $R=dC-CH$ are assumed invariant they are annihilated by $s$,
from which it follows that
\begin{eqnarray}
sB &=& d\Delta^{(NS)} \nonumber\\
sC &=& d\Delta^{(R)} -\Delta^{(R)} H \, ,
\label{symd}
\end{eqnarray}
where $\Delta^{(NS)}$ is a ghost-valued superspace 1-form and $\Delta^{(R)}$ is 
a formal sum of ghost-valued superspace forms of all (relevant) degrees.
Alternatively, these quantities may be viewed as 1-forms on the isometry group
manifold with values in the exterior algebra on superspace. Either way, we see
that if the BRST transformations of the worldvolume gauge fields $V$ and $A$ are
chosen to be
\begin{equation}
sV= \Delta^{(NS)}, \qquad sA = e^{\cal F} \Delta^{(R)}\, ,
\label{symg}
\end{equation}
then we have
\begin{equation}
s{\cal F}=0, \qquad s{\cal G}=0\, .
\label{symga}
\end{equation}
In other words, {\it if the superspace background is such that the Lie
derivative of each background tensor vanishes then transformations of the
independent worldvolume gauge  potentials $V$ and $A$ can be chosen such that
their respective field strengths ${\cal F}$ and ${\cal G}$ are invariant}, from
which it follows that the super D-brane action is invariant. 

Having established invariance under background isometries, the next step is to
compute the algebra of these symmetry transformations. Since the Lagrangian is
invariant, and not just the action, this is equivalent to a computation of the
algebra of Noether charges in the Hamiltonian formulation. Since $B$ and $C$
are superfields we have the identities $s^2B\equiv 0$ and $s^2C\equiv 0$, which
imply that
\begin{equation}
d{\cal A}^{(NS)} =0, \qquad d{\cal A}^{(R)} =0,
\label{syme}
\end{equation}
where
\begin{equation}
{\cal A}^{(NS)} = s\Delta^{(NS)},\qquad 
{\cal A}^{(R)} = e^{\cal F} s\Delta^{(R)}.
\label{symf}
\end{equation}
When these closed superspace forms are exact the transformations of $V$ and
$A$ can be removed by gauge transformations and are therefore trivial. Thus 
${\cal A}^{(NS)}$ and ${\cal A}^{(R)}$ may be viewed as 2-forms on
the isometry group manifold with values in cohomology classes of superspace.

Because $V$ and $A$ are not pullbacks of superspace forms they need not be
annihilated by $s^2$. In fact,
\begin{equation}
s^2 V = {\cal A}^{(NS)} \qquad s^2 A={\cal A}^{(R)}\, .
\label{symh}
\end{equation} 
Note that the closure of ${\cal A}^{(NS)}$ and ${\cal A}^{(R)}$ ensures
that $s^2$ annihilates ${\cal F}$ and ${\cal G}$ (as is, of course, 
guaranteed by the construction). The closed forms defined by $s^2V$ and $s^2A$
determine the topological charges appearing as central charges in the algebra of
isometries. They may be calculated explicitly for any particular background. An
equivalent explicit calculation for a flat superspace background has been
carried out in \cite{hammer}.

\section{The super D-brane hamiltonian}

In passing to the phase-space form of the D-brane action it is useful to
first consider the null super D-brane, for which the Lagrangian is 
\cite{lindstrom},
\begin{equation}
L = {1\over 2v} e^{-2\phi}\det (g_{ij}+ \CF_{ij})\, .
\label{hama}
\end{equation}
This is obtained by setting $G=0$ in (\ref{onea}). We can rewrite this as
\begin{equation}
L = {1\over 2v} e^{-2\phi} \det (g_{\ua\ub}+\CF_{\ua\ub}) \big\{ g_{tt} -
(g_{t\ua} +
\CF_{t\ua}) [(g+\CF)^{-1}]^{\ua\ub} (g_{t\ub} + \CF_{t\ub}) \big\}
\label{hamb}
\end{equation}
where we have set $i=(t,\ua_1,\ua_2,\dots,\ua_p)$, i.e. an underlined lower-case
latin index is a `worldspace' index. The matrix $(g+\CF)^{-1}$, with
only worldspace components, is the inverse of the matrix with (only
worldspace) components $(g_{\ua\ub}+\CF_{\ua\ub})$. This Lagrangian can be
further rewritten as
\begin{equation}
L = {1\over 2v} e^{-2\phi}\det (g_{\ua\ub}+\CF_{\ua\ub}) \big[ K_{tt} - K_{t\ua}
(K^{-1})^{\ua\ub} K_{t\ub}\big]
\label{hamc}
\end{equation}
where
\begin{equation}
K_{ij} = g_{ij} + \CF_{i\ua}(g^{-1})^{\ua\ub}\CF_{j\ub}\, .
\label{hamd}
\end{equation}
Here, both $K^{-1}$ and $g^{-1}$ are to understood as being the inverses of
the matrices with only worldspace components, i.e. of $K_{\ua\ub}$ and
$g_{\ua\ub}$ respectively. If we now define
\begin{equation}
\lambda = ve^{2\phi}/\det (g_{\ua\ub}+\CF_{\ua\ub})
\label{hame}
\end{equation}
then we have
\begin{equation}
L= {1\over 2\lambda}\big[ K_{tt} - K_{t\ua} (K^{-1})^{\ua\ub}K_{t\ub}\big]\, .
\label{hamf}
\end{equation}
An equivalent Lagrangian is
\begin{equation}
L = \tilde P\cdot \Pi_t + \tilde E^\ua \CF_{t\ua} - 
s^\ua (\tilde P\cdot \Pi_\ua + \tilde E^\ub \CF_{\ua\ub})
- {1\over2} \lambda (\tilde P^2 + \tilde E^\ua \tilde E^\ub g_{\ua\ub})
\label{hamg}
\end{equation}
where 
\begin{equation}
(\Pi_i)^a \equiv E_i{}^a \, .
\label{hamh}
\end{equation}
To establish the equivalence one first eliminates the new auxiliary variables
$\tilde P_a$ and $\tilde E^\ua$ to obtain the Lagrangian
\begin{equation}
L= {1\over 2\lambda} [ K_{tt} - 2s^\ua K_{t\ua} + s^\ua s^\ub K_{\ua\ub} ]
\label{hami}
\end{equation}
Elimination of $s^\ua$ then yields (\ref{hamf}).

The Lagrangian (\ref{hamg}) is a convenient `half-way house' on the way to the
true phase-space form of the null D-brane action (the variable $\tilde P$ is not
the momentum conjugate to $X$, for example, although it is closely related to 
it). However, rather than proceed with the null D-brane it is convenient to now
pass to the full D-brane action. We first note that the worldvolume (p+1)-form
$G_{(p+1)}$ in the formal sum $G$ can be written as
\begin{equation}
G_{(p+1)}= dt\, (G_t)_{(p)} \, ,
\label{hamj}
\end{equation}
where $(G_t)$ is a p-form on the p-dimensional worldspace. It is the p-form in
the formal sum
\begin{equation} 
G_t = \dot A - dA_t - \dot Z^M C_M e^\CF - \CF_t (Ce^\CF)
\label{hamk}
\end{equation}
where $A$ is now a formal sum of {\sl worldspace} forms, as is $A_t$, and
$d$ is now to be understood as an exterior derivative on worldspace.
Similarly, $C_M$ is the restriction to worldspace of the formal sum 
$i_MC$, where $i_M$ denotes contraction with the vector superfield
$\partial/\partial Z^M$. Finally, $\CF_t = d\sigma^\ua \CF_{t\ua}$ is a
worldspace 1-form. Specifically,
\begin{equation}
\CF_t = \dot V - dV_t - \dot Z^M B_M\, ,
\label{haml}
\end{equation}
where $B_M$ is the restriction to worldspace of the 1-form $i_MB$, 
so that
\begin{equation}
G_t = \dot A - dA_t - \dot V (Ce^\CF)
- \dot Z^M [ C_M e^\CF - B_M (Ce^\CF)] 
+ (dV_t) (Ce^\CF)\, .
\label{hamm}
\end{equation}
Using these results we can now write down a `half-way house' version of the {\sl
full} super D-p-brane Lagrangian (\ref{onea}), involving $G_t$. This is
\begin{eqnarray}
L &=& \tilde P\cdot \Pi_t + \tilde E^\ua \CF_{t\ua} + T*(G_t)_{(p)}
- s^\ua (\tilde P\cdot \Pi_\ua + \tilde E^\ub \CF_{\ua\ub})\nonumber\\
&&- {1\over2} \lambda [\tilde P^2 + \tilde E^\ua \tilde E^\ub g_{\ua\ub} 
+ T^2 e^{-2\phi} \det (g_{\ua\ub}+\CF_{\ua\ub})]\, 
\label{hamja}
\end{eqnarray}
where $*$ indicates the {\sl worldspace} Hodge dual. The equivalence can be 
established as before by elimination of the variables $\tilde P,
\tilde E^\ua$ and  $T$ followed by the redefinition (\ref{hame}) and elimination
of
$s^\ua$: the Lagrangian (\ref{onea}) is then recovered. 

To obtain this action in canonical form we proceed as follows. Omitting total
derivatives, and defining the (p-1)-form
\begin{equation}
U_{(p-1)} = (A_t)_{(p-1)} + (-1)^p V_t (Ce^\CF)_{(p-1)}\, ,
\label{hamma}
\end{equation}
we have
\begin{eqnarray}
T(G_t)_{(p)} &=\dot A_{(p)} T + U_{(p-1)} dT - \dot V T (Ce^\CF)_{(p-1)}
\nonumber\\
& -\dot Z^M T[C_M e^\CF - B_M(Ce^\CF)]_{(p)} + (-1)^p V_t T (Re^\CF)_{(p)} 
\label{hammab}
\end{eqnarray}
where $R= dC -CH$ is the field strength for $C$\footnote{As before we use the
standard superspace convention that exterior differentiation `starts from the
right'.}. Thus
\begin{equation}
T*(G_t)_{(p)} = \dot \phi T + \phi^\ua \partial_\ua T - T\dot V_{\ua} {\cal
C}^{\ua} - T\dot Z^M [{\cal C}_M - (B_M)_\ua {\cal C}^\ua] + (-1)^pT V_t {\cal R}
\label{hamn}
\end{equation}
where we have defined
\begin{equation}
\phi = *A_{(p)}\,,\qquad \phi^\ua = (*U_{(p-1)})^\ua\,,
\label{hamna}
\end{equation}
and
\begin{equation}
{\cal C}^\ua = [*(Ce^\CF)_{(p-1)}]^\ua\, \qquad 
{\cal C}_M = * (C_M e^\CF)_{(p)}\, \qquad
{\cal R} = *(Re^\CF)_{(p)} \, .
\label{hamo}
\end{equation}

Thus (\ref{hamja}) can now be rewritten as
\begin{equation}
L= \dot Z^M P_M + \dot V_\ua E^\ua + \dot\phi T - H\, ,
\label{hamp}
\end{equation}
where the hamiltonian $H$ is a sum of constraints imposed by Lagrange
multipliers, and
\begin{eqnarray}
P_M &=& E_M{}^a \tilde P_a - \tilde E^\ua (B_M)_\ua - 
T[{\cal C}_M - (B_M)_\ua {\cal C}^\ua]\, , \nonumber\\
E^\ua &=& \tilde E^\ua - T {\cal C}^\ua\, .
\label{hamq}
\end{eqnarray}
These equations imply that
\begin{eqnarray}
\tilde P_a &=& E_a{}^M \big(P_M + E^\ua (B_M)_\ua + T{\cal C}_M\big)\, ,
\nonumber\\
\tilde E^\ua &=& E^\ua + T{\cal C}^\ua\, . 
\label{hamr}
\end{eqnarray}
Since $E_\mu{}^\alpha$ is invertible, the remaining information contained in
(\ref{hamr}) is the constraint
\begin{equation}
E_\alpha{}^M\big(P_M + E^\ua (B_M)_\ua + T{\cal C}_M\big)=0\, ,
\label{hams}
\end{equation}
which can be incorporated as a constraint in $H$ imposed by a new fermionic
Lagrange multiplier $\zeta$. Thus 
\begin{equation}
H = \phi^\ua {\cal T}_\ua + V_t {\cal G} + s^\ua {\cal H}_\ua + 
\lambda {\cal H} +\zeta^\alpha {\cal S}_\alpha\, ,
\label{hamt}
\end{equation}
where
\begin{eqnarray}
{\cal T}_\ua &=& -\partial_\ua T \nonumber\\
{\cal G} &=& - \partial_\ua {\tilde E}^\ua + (-1)^{p+1} T {\cal R}\nonumber\\
{\cal H}_\ua &=& {\tilde P}\cdot \Pi_\ua + {\tilde E}^\ub {\cal
F}_{\ua\ub}\nonumber\\ 
{\cal H} &=& {1\over2}\big[
{\tilde P}^2 + {\tilde E}^\ua {\tilde E}^{\ub} g_{\ua\ub} + T^2 e^{-2\phi}
{\rm det}(g+{\cal F}) \big]\nonumber\\
{\cal S}_\alpha &=& E_\alpha{}^M\big[P_M + E^\ua (B_M)_\ua + T{\cal C}_M\big]
\label{hamu}
\end{eqnarray}
with $\tilde P_a$ and $\tilde E^\ua$ given by (\ref{hamr}). 
The constraint functions ${\cal T}_\ua$, ${\cal G}$, ${\cal H}_\ua$ and ${\cal
H}$ are `first class' (in Dirac's terminology) and generate the p-form gauge
transformations, BI gauge transformations, worldspace diffeomorphisms and time
translations, respectively. The fermionic constraint functions ${\cal S}$ are
half first class and half second class; the first class constraints generate
$\kappa$-symmetry transformations.

\section{Discussion}

In this paper we have presented a new formulation of the super D-p-brane action
in which the tension appears as an integration constant in the equation of
motion of a new p-form gauge potential. In this form of the action the full
centrally-extended supertranslation algebra is realized as the `naive' algebra
of transformations of worldvolume fields. In the standard form, only the NS
charges are `naive' while the RR charges arise from the presence of the WZ
term. In the zero tension limit the RR charges vanish and one is left with the
`naive' algebra of the null super D-brane. The latter does not coincide with
the standard supersymmetry algebra (in contrast to, for example, the null D=11
supermembrane) because the BI field does not decouple in the null limit.

The results just summarized provide a clear understanding of the origin of the
various p-form charges in the supertranslation algebra of the super
D-p-brane. One of the original motivations for this work was to obtain a
similar understanding of the recently determined central charge structure of the
M5-brane \cite{sorokin}. In that case the WZ term was found to be responsible for
the full 5-form charge but for only half of the 2-form charge. The remaining
half is explained by the fact that, because of the worldvolume 2-form gauge
potential, the `naive' supersymmetry algebra differs from the standard one. The
D-brane results reported here are similar but with the additional feature that
one can exhibit the `naive' algebra as the algebra in the null limit, thereby
isolating the naive (NS) and WZ (RR) contributions. In the M5-brane case there
is apparently no way to achieve this separation: our attempts to define a null
limit of the M5-brane (in the `temporal gauge') simply led to the standard null
super 5-brane in which the two-form potential is absent\footnote{There is no
conflict with supersymmetry here because, as pointed out in \cite{blt}, spacetime
supersymmetry  and $\kappa$ supersymmetry do {\it not} imply worldvolume
supersymmetry for null branes.}. In view of this, one might try instead to unify
the contributions to the central charge structure of the M5-brane supersymmetry
algebra by realizing the full algebra as the naive supersymmetry algebra on
worldvolume fields, as is achieved for D-branes via the new scale-invariant
action presented here. It is not yet clear to us whether this is possible. As
pointed out in \cite{pktb}, a p-form worldvolume gauge potential is natural for
objects that may, like D-branes, have boundaries on other branes, but the
M5-brane is always closed. A 5-form gauge potential has been successfully
introduced in a recent reformulation of the M5-brane action \cite{swedesb}, but
as the self-duality constraint is incorporated at the level of the field
equations this action could not be used to extract the central charge structure
in the manner envisaged here. We shall return to the M5-brane in a future
publication \cite{BSTb}.

We have also shown that the new super D-brane Lagrangian is invariant under all
isometries of the supergravity background, provided that the worldvolume
gauge fields are taken to transform appropriately; this implies that the `old'
super D-brane Lagrangian is invariant up to a total derivative. One advantage
of the new Lagrangian is that its invariance means that the full symmetry
algebra, with any possible central extensions, may be deduced from the algebra
of transformations of the worldvolume fields. Central extensions of the algebra
of Killing vector superfields are coded in the action of a BRST operator $s$ on
the worldvolume gauge fields: one finds that $s^2$ generates gauge
transformations of these fields with parameters determined by closed
superspace forms. 

A case of current interest to which this analysis is applicable is the super
D3-brane in the maximally supersymmetric $adS_5\times S^5$ IIB background,
which has isometry supergroup $SU(2,2|4)$. An appropriate embedding of a super
D3-brane worldvolume in this background has been shown to result in an
interacting worldvolume field theory in which the $SU(2,2)$ subgroup acts as a
non-linearly realized conformal symmetry on the bosonic fields \cite{kaletal}.
The results obtained here show that the full action is  invariant under the full
$SU(2,2|4)$ isometry supergroup, which can be interpreted as a non-linearly
realized superconformal symmetry. More precisely, we have shown that the action
is invariant under transformations that close to $SU(2,2|4)$  on the worldvolume
scalar (and, implicitly, spinor) fields. Closure on the worldvolume gauge fields
might, in principle, require the introduction of additional charges that are
central with respect to the linearly realized subgroup of $SU(2,2|4)$, which is
$(3+1)$-dimensional super-Poincar{\'e} and its $SU(4)$ group of automorphisms.
Specifically, one might expect the D3-brane to be associated with a 3-form
charge in the adS superalgbra. From the perspective of $adS_5$ the space
components of this 3-form (which are naturally associated with a 3-brane) would
be dual to a time component of a dual 2-form, but there is already a charge of
this type in the adS algebra: it is the generator of boosts in the space
direction orthogonal to the 3-brane. That this must be the 3-brane charge can be
seen from the fact that the $adS_5$ spacetime can itself be viewed as a
$p$-brane \cite{pope}, for  which the only candidate charge is the one already in
the adS algebra. There are therefore no {\sl new} charges in the symmetry
algebra relative to the algebra of isometries, so that the symmetry
group of the super D3-brane is just the superconformal group $SU(2,2|4)$.

\end{document}